\documentstyle[12pt,aaspp4,psfig]{article}

\newcommand{\kms}          {\mbox{${\rm km~s^{-1}}$}}
\newcommand{\cc}           {\mbox{${\rm cm^{-3}}$}}

\newcommand{\e}            {\mbox{$^{-1}$}}
\newcommand{\ee}           {\mbox{$^{-2}$}}

\newcommand{\simgt}        {\gtrsim}
\newcommand{\simlt}        {\lesssim}

\newcommand{\Kkms}         {\mbox{${\rm K\,km\,s}^{-1}$}}

\def\cm2{\mbox{${\rm cm^{-2}}$}}
\def\h2{\mbox{${\rm H}_2$}}
\def\nh2{\mbox{$n_{\rm H_2}$}}
\def\Nh2{\mbox{$N_{{\rm H}_2}$}}
\def\Mh2{\mbox{$M_{{\rm H}_2}$}}
\def\n2hp{\mbox{N$_2$H$^+$}}
\def\c34s{\mbox{C$^{34}$S}}
\def\fs{\hbox{$.\!\!^{\rm s}$}}
\def\vin{\mbox{$v_{\rm in}$}}

\begin{document}

\title{A HIGH RESOLUTION STUDY OF THE SLOWLY CONTRACTING, STARLESS CORE L1544}
\author{Jonathan P. Williams, Philip C. Myers,
David J. Wilner, and James Di Francesco}
\affil{Harvard--Smithsonian Center for Astrophysics, 60 Garden Street,
Cambridge, MA 02138}

\begin{abstract}
\rightskip = 0pt
We present interferometric observations of \n2hp(1--0)
in the starless, dense core L1544 in Taurus.
Red-shifted self-absorption, indicative of inward motions,
is found toward the center of an elongated core.
The data are fit by a non-spherical model consisting of two
isothermal, rotating, centrally condensed layers. Through a
hybrid global-individual fit to the spectra,
we map the variation of infall speed at scales $\sim 1400$~AU
and find values $\sim 0.08$~\kms\ around the core center.
The inward motions are small in comparison to thermal,
rotational, and gravitational speeds but are large enough
to suggest that L1544 is very close to forming a star.
\end{abstract}


\rightskip = 0pt
\section{Introduction}
The very first stage in models of the star formation process
is the gravitational collapse of a {\it starless} dense molecular core
(e.g., Shu, Adams, \& Lizano 1987)
but this stage is relatively short-lived, and therefore rare.
One of the few known examples is the dense
core, L1544, in Taurus which shows no evidence of an embedded young
stellar object but appears to have extended inward motions
(Tafalla et al. 1998; hereafter T98).
This core is therefore an excellent testing ground for
theories of low mass, isolated, star formation.

T98 presented single-dish molecular line observations
and analyzed the large scale structure and dynamics of L1544.
They find that lines of CO and its isotopes are single peaked
and trace an outer envelope, but higher-dipole-moment lines of 
H$_2$CO, CS, and \n2hp\ are strongly self-absorbed
and trace a central core. The self-absorption is predominantly
red-shifted as expected for inward motions
(e.g., Myers et al. 1996; hereafter M96)
although there is also a small region of blue-shifted reversals
toward the south. The inferred inward speeds range from 
$0.10$~\kms\ in the northeast to $-0.03$~\kms\ (relative outward motions)
in the region of reversals in the south.
T98 note that the dynamics differ
from several theoretical predictions for core collapse:
the infall speeds are too large for ambipolar diffusion
(Ciolek \& Mouschovias 1995);
the spatial extent of infall is too large for inside-out collapse
(Shu 1977) at such an early stage;
and the density is too highly centrally condensed to be
due to collapse from a uniform density configuration (Larson 1969).

In order to measure the structure and dynamics in greater detail,
we made high resolution observations of \n2hp(1--0).
This line was chosen because the T98 observations show a compact,
centrally condensed core with double-peaked spectra. These observations
are the first kinematic measurements of a starless core on the size scale 
$\sim 2000$~AU ($\sim 0.01$~pc). Our main finding is that the maximum
infall speed is (surprisingly) essentially the same,
$\sim 0.1$~km~s$^{-1}$, as on much larger scales.

\section{Observations}
\label{sec:obs}
L1544 was observed with the ten-antenna Berkeley-Illinois-Maryland array
\footnotemark\footnotetext{Operated by the University of California at
Berkeley, the University of Illinois, and the University of Maryland,
with support from the National Science Foundation}
(BIMA) in its compact C configuration for two 8 hour tracks
on November 4 and 5, 1997.
Projected baselines ranged from 1.9 to 25.1 k$\lambda$.
The phase center was
$\alpha(2000)=5^{\rm h}04^{\rm m}16\fs 62,
~\delta(2000)=25^\circ 10'47\farcs 8$.
Amplitude and phase were calibrated using 3 minute observations of
0530+135 interleaved with each 25 minute integration on source.
The passband shape was determined from observing the bright quasar 3C\,454.3.
No suitable planets were available for flux calibration
and we therefore assumed a flux density of 2.4~Jy for
0530+135 based on flux density measurements taken within a month of
these observations. From the scatter in these measurements, the flux
density scale is accurate to 30\%.

The digital correlator was configured with 512 channel windows
at a bandwidth of 6.25~MHz ($0.04$~\kms\ velocity resolution per channel)
centered on the 7 hyperfine components of \n2hp(1--0)
(Caselli, Myers, \& Thaddeus 1995) in the lower sideband
and \c34s(2--1) in the upper sideband.
Eight 32 channel windows at a bandwidth of 100~MHz.
were used to measure continuum radiation.
Data were reduced with the MIRIAD package using standard procedures.
The data sets from each day were calibrated separately and then transformed
together using natural weighting. The resulting ``dirty'' map
was cleaned and restored with a gaussian beam of FWHM size
$14\farcs 8\times 6\farcs 6$ at position angle $-1^\circ$.
The continuum map shows a 5~mJy peak, corresponding to $5\,\sigma$,
at offset position, $\Delta\alpha=4'', \Delta\delta=12''$,
but this has not been verified by the IRAM Plateau de Bure interferometer
(J.-F. Panis, personal communication) and may be due to contamination of
low level line emission from species such as CH$_3$OH in the wide windows.
We also searched for continuum emission at 3.6~cm
from a very young protostellar wind using the VLA
in B-configuration on July 5, 1998.
No emission was detected to a $3\,\sigma$ level of 39~$\mu$Jy
which is considerably less than the outflows detected by
Anglada (1995) and, extrapolating his data, implies an upper limit
to the bolometric luminosity of any protostar in L1544 of $0.03~L_\odot$.

The \n2hp\ lines were detected with high signal-to-noise but
no emission was detected in the other spectral window implying a
$3\,\sigma$ upper limit of 0.25~\Kkms\ for the \c34s\ emission
integrated over velocities 6.8 to 7.6~\kms.
A map of the velocity integrated \n2hp\ emission for the isolated
hyperfine component F$_1$F$=01-12$ is displayed in Figure~1.
There is a single, compact, elongated structure
with FWHM size 7000~AU$\times$3000~AU.
Double-peaked spectra are found toward the map center as in T98,
but before analyzing these profiles it is necessary to account for
the effect of the missing zero-spacing flux (see Gueth et al. 1997).
Therefore, we combined the single-dish IRAM 30~m data of T98 with the
BIMA map: the single-dish data were shifted by $13''$ to the south and
then scaled to match the interferometer data in the region of visibility
overlap (6~m to 30~m) using a best fit gain $4.8$~Jy~K\e.
The resolution of the resulting data set is the same as for the
interferometer data alone but a circular restoring beam of FWHM size $10''$
(having approximately the same beam area) was used to
create a regular grid of spectra for analysis.

\section{Analysis}
\label{sec:analysis}

The combined IRAM-BIMA spectra are also double-peaked toward the map center
and we attribute this to self-absorption for the reasons outlined in T98.
The red peak is stronger than the blue as expected for a contracting core
but the contrast between the two is not as great as
in the interferometer map alone which implies that the absorbing region is
more spatially extended than the emitting region.

To examine the structure and dynamics of the core in further detail,
we fit the data with a simple model of a collapsing core.
The self-absorption implies a decreasing excitation
temperature gradient away from the core center toward the observer.
Many previous models of core collapse (e.g., Zhou 1995)
have assumed a spherical gas distribution, but Figure~1 shows that the core
is highly elongated with an axial ratio $\sim 0.4$,
so such models are inappropriate in this case.
Here, we allow for variations in parameters parallel and perpendicular
to a major axis in the plane of the sky, and consider two isothermal layers,
each of constant thickness, but at different densities so as
to provide the excitation temperature inversion.
We were motivated to consider this heuristic model because it greatly
simplifies the treatment of the radiative transfer yet fits
individual spectra quite well. It allows us to infer the variation of
density in the plane of the sky and the magnitude and spatial distribution
of the relative motion between the two layers (the infall speed).

The line brightness temperature is given by M96 equation (2)
and depends on the excitation temperature and optical depth of each layer.
The excitation temperatures are derived using a two level approximation as
in M96 equation (8a,b) and the peak optical depth of the 101-012 transition is
$$\tau_{{\rm pk},i}=
  \left({{Xn_i\Delta z_i}
  \over{4.9\times 10^{12}~{\rm cm}^{-2}}}\right)
  \left({{1-e^{-4.5/T_{{\rm ex},i}}}
  \over{\sigma T_{{\rm ex},i}}}\right),
\eqno(1)$$
for $\tau_i=\tau_{{\rm pk},i}{\rm exp}[-(v-v_i)^2/2\sigma^2]$,
where $v_i$ is the systemic velocity and $\sigma$ is the velocity dispersion.
Here, the index $i=f$ or $r$ for the front or rear layer
respectively, and $X$ is the abundance of N$_2$H$^+$ relative to H$_2$,
$n_i$ is the density of H$_2$, $\Delta z_i$ the layer thickness,
and $T_{{\rm ex},i}$ the excitation temperature.

This formalism allows each spectrum to be fit very well
but requires at least 6 independent parameters for each spectrum (M96).
Instead, we present a global fit to the data that describes the core
structure and dynamics with the addition of fewer parameters.
The route to a global fit was guided by the individual fits;
these showed that the velocity dispersion did not vary greatly but the
excitation temperature and optical depth both increased sharply toward the
map peak.
The kinetic temperature is constrained to the range $T_{\rm k}=11-13.5$~K
by the CO observations of T98 and is insufficient to account for the range
of excitation temperature, implying that the density must increase toward 
the map center. We therefore fixed $T_{\rm K}$ and $\sigma$ and chose the
functional form,
$$f(x,y)=1+\left({x-x_{\rm pk}}\over{\Delta x}\right)^2
          +\left({y-y_{\rm pk}}\over{\Delta y}\right)^2,
\eqno(2)$$
to describe the spatial variation of the density,
$n_i(x,y)=n_{{\rm pk},i}/f(x,y)$,
where $x-y$ defines an orthogonal coordinate system rotated with
respect to the $\alpha-\delta$ observational frame by an angle $\theta$
(measured anti-clockwise from north),
chosen so that $\Delta x\geq\Delta y$ (i.e., $x$ is the major axis).
The kinetic temperature and velocity dispersion are then held
constant, both spatially within a layer and from layer to layer.

The individual fits (and also channel maps) show that there is a
significant velocity gradient that is approximately uniform and along
the major axis and therefore we set
$$v_r(x)=v_0+(dv/dx)(x-x_{\rm pk}),~~~
v_f(x,y)=v_r(x)+v_{\rm in}(x,y),
\eqno(3)$$
where $v_0$ is a constant velocity offset,
$dv/dx$ is a constant velocity gradient ($dv/dy\equiv 0$),
and $\vin=v_f-v_r$ is the velocity difference
between the front and rear layers (the infall speed).

In addition to \vin, there are, formally,
a total of 14 parameters in the final model:
$x_{\rm pk}, y_{\rm pk}, \theta$ to define the coordinate system;
$n_{\rm pk,f}, \Delta z_f, n_{\rm pk,r}, \Delta z_r,
\Delta x, \Delta y$ to describe the densities in each layer;
the velocity parameters, $v_0$ and $dv/dx$;
and the constants, $T_{\rm k}, \sigma$, and $X$.
This is far fewer than the $>$200 parameters needed
to fit the spectra in Figure~2 using 6 parameters per spectrum.
Furthermore, the parameter space is quite tightly constrained
by the maps showing the approximate center, angle, and velocity
gradient of the core, and the estimates of kinetic temperature, sizes,
and linewidths that were determined by T98.
Also, not all the parameters are independent;
all values of $\Delta z_i$ and $X$ with the
same product result in the same model output.
The principal parameters that were varied were the six density
parameters and the infall speed.

We proceeded by assuming a constant value for $v_{\rm in}(x,y)$,
and searched for the best global fit by least squares minimization of the
difference between the model and observed spectra for a grid of
$5\times 7$ spectra at $10''$ spacing about the core center (Figure~2).
This showed that any viable fit must satisfy the following conditions:
(1) $n_{\rm pk,r}\simgt n_{\rm cr}\gg n_{\rm pk,f}$
and $\Delta z_r\ll\Delta z_f$
where $n_{\rm cr}$ is the critical density;
the excitation temperature must be high in the rear
which emits strongly, and low in the front which emits weakly
but has a large line-of-sight thickness and absorbs strongly.
(2) $\Delta x, \Delta y$ $<$ (core size); the excitation temperature and
optical depth must decrease rapidly away from the map center.
(3) $dv/dx\neq 0$; a constant systemic velocity is a very poor fit and
a velocity gradient is required.
(4) $\vin >0$; the front layer must be moving toward the
rear layer to account for the asymmetry of the self-absorption.
In fact, many of the spectra do not show two distinct peaks
but rather a bright red peak and a blue shoulder indicating that the
velocity of the absorption must be shifted relative to the emission
by an amount approximately equal to the velocity dispersion,
$\vin\sim\sigma$.

Model spectra are plotted with the observations in Figure~2 for the 
best fit parameter values listed in Table~1.  We chose a high value 
for the abundance, $X=1\times 10^{-9}$, so that the front layer (with 
peak density $n_{\rm pk,f}=10^4$~\cc) has a symmetrical size, $2\Delta 
z_f=0.4$~pc, that approximately matches the region of C$^{18}$O emission 
in T98.  The infall speed was determined for each spectrum individually 
by least squares minimization after the other parameters were fixed.
This hybrid global-individual approach was chosen to emphasize
the spatial variation of \vin.
However, its determination requires high signal-to-noise ratios
and optical depth, and it could only be reliably measured around
the map center where there are bright lines and strong self-absorption.
We find that the best fit has \vin\ increasing toward the core center
and along the major axis, but because the optical depth falls off rapidly,
the significance of this fit is only marginally better
than a fit with constant $\vin=0.075$~\kms.
The difference between the model spectra and the data has a mean of zero
and standard deviation $0.12$~K per 0.04~\kms\ channel,
about 50\% greater than the rms noise in the spectra.
The spatial variation of column density and infall speed are
displayed along with the spectra in Figure~2.

\section{Discussion}

The observations reported here are the first to define the inward motions
in a starless core on the size scale of star formation, i.e. on a size scale
which encloses about a stellar mass.  We therefore comment here on the model
parameters, the inferred core motions and their physical basis, and the 
evolutionary status of L1544.

Our model properties agree with four independent constraints.
We find agreement between the size scale, $\Delta z_f$, as discussed above,
and the kinetic temperature, $T_{\rm k}=12$~K, from the T98 CO observations.
The core mass can be derived by integrating the density profile out to a
distance equal to the spatial thickness of each layer:
$M_f=0.2~M_\odot, M_r=0.4~M_\odot$ implying a total
symmetric mass $2(M_f+M_r)=1.2~M_\odot$, which is comparable
to estimates of the dense gas mass discussed in T98.
The density profile and peak column density are consistent with
dust continuum measurements of the structure of a similar prestellar
core, L1689B, in the $\rho$ Ophiuchus cloud
(Andr\'{e}, Ward-Thompson, \& Motte 1996).
We have also fit the data with the exponent in equation (2) varying
from 1.8 to 2.5 (with corresponding changes in $\Delta x$ and $\Delta y$)
without significantly increasing the least squares difference.

The core dynamics are of greatest interest.
The dispersion, $\sigma=(\sigma_{\rm T}^2+\sigma_{\rm NT}^2)^{1/2}$,
is the quadrature sum of a thermal component,
$\sigma_{\rm T}=(kT_{\rm k}/\mu)^{1/2}$, where $k$ is the Boltzmann constant
and $\mu$ is the molecular mass, and a non-thermal component,
$\sigma_{\rm NT}$, which has a best fit value of $0.085$~\kms,
similar to that measured for \c34s\ by T98.
For $T_{\rm k}=12$~K, $\sigma_{\rm T}({\rm H}_2)=0.22$~\kms, 
much greater than $\sigma_{\rm NT}$ and comparable to
gravitational speeds $\sim (GM_r/\Delta z_r)^{1/2}=0.27$~\kms.
Therefore, the turbulent pressure support is negligible.
However, the non-spherical core shape implies that it cannot be
supported entirely by thermal pressure and there must also be an
anisotropic force resisting self-gravity. The velocity gradient along
the major axis, $dv/dx=3.8$~\kms~pc\e,
implies a change in velocity $\Delta v_{\rm rot}=0.13/{\rm sin} i$~\kms\
over the 7000~AU major axis FWHM core diameter, where $i$ is the 
inclination of the gradient to our line of sight.  If the non-sphericity 
is due to rotation, however, the aspect ratio constrains ${\rm cos} i<0.4$, 
so $\Delta v_{\rm rot}<0.14~\kms\ll\sigma_{\rm T}({\rm H}_2)$ and we 
conclude that rotation is dynamically insignificant.  This leaves magnetic 
fields as the most likely explanation for the extended core shape. The 
observed aspect ratio implies approximate equipartition between thermal 
and {\it static} magnetic pressure support (Li \& Shu 1996), corresponding 
to a field strength $B\simeq 30~\mu$G at number densities $\simeq 10^5$~\cc.

The thermal, gravitational, and rotational speeds are all greater
than the maximum infall speed. We have determined a range,
$\vin\simeq 0.02$ to $0.09$~\kms, that is similar to
the values derived from the CS/\c34s\ analysis by T98
even though the maps are quite different in spatial scale.
The interferometer map presented here has finer resolution,
but covers a much smaller spatial extent than the T98 CS map.
Comparing the two maps, the high resolution observations indicate a sharp
increase in density, consistent with free-fall collapse, but no corresponding
increase in infall speed.

A possible explanation is that starless cores, such as L1544,
are predicted to evolve through a series of quasi-static
equilibria (see discussion in Li \& Shu 1996).  For example,
neutrals in an outer envelope diffuse through magnetic field lines
onto a central core which is supported by thermal pressure.  The 
piling up of material on the central core causes its density to
increase and its (Jeans) size to decrease.  The mass-to-flux ratio 
increases until the collapse becomes supercritical and magnetic field 
lines are dragged into the core which remains (quasi-)static.  The 
infall velocities remain sub-sonic except for a small region at the 
envelope-static core boundary at very late times (Li 1998; Basu \& 
Mouschovias 1994).  As with T98, however, the quantitative comparison 
shows discrepancies.  For example, both Ciolek \& Mouschovias (1995) 
and Basu \& Mouschovias (1994) predict smaller infall velocities than
measured here, i.e., $\simlt 0.02$~\kms, at size scales of 0.02~pc and 
densities of $4\times 10^5$~\cc.  The large scale ($\sim 0.1$~pc) inward 
motions observed by T98, possibly driven by turbulent pressure gradients 
(Myers \& Lazarian 1998) and not included in the above models, probably 
increase the infall speed at all size and density scales.

\n2hp\ is an ion and therefore moves only along, or with,
the magnetic field lines. Since the ions and neutrals are moving
together at a speed much greater than their relative drift speed,
either we are looking down along the field lines,
or the magnetic field lines are being dragged in along with the gas.
The geometry suggests that the field lines lie more closely perpendicular
than parallel to our line of sight, and therefore the second possibility,
supercritical collapse, seems more reasonable.
In this case, L1544 may be very close to forming a star:
Li (1998) shows that, for a spherical cloud, both the ion and
neutral infall speeds are very small for times $t\simlt 15\tau_{\rm ff}$,
where $\tau_{\rm ff}\simeq 0.4$~Myr is the free-fall timescale,
but then rapidly increase for $t\simgt 19\tau_{\rm ff}$
and the core forms a star by $20\tau_{\rm ff}$.
There may be many cores that have begun collapsing in Taurus,
but these calculations suggest that only those that are more than 95\% of
the way there, such as L1544 appears to be, may show detectable inward motions.
Observational verification, however, awaits the statistics from surveys
for inward motions in starless cores.

\acknowledgments
Discussions with Zhi-Yun Li and Shantanu Basu are gratefully acknowledged.
This research was partially supported by NASA Origins grant NAGW-3401.

\clearpage

\clearpage

\begin{figure}[htpb]
\centerline{\psfig{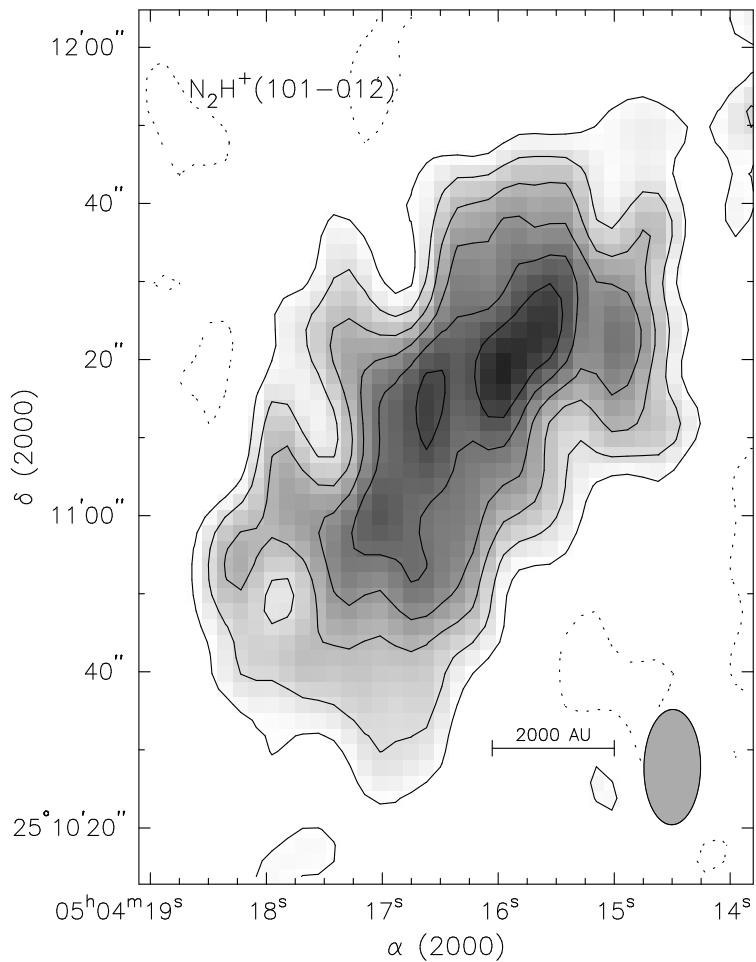}}
\caption{BIMA map of the integrated intensity map of \n2hp(101-012) emission
in L1544. The emission has been summed over velocity channels from
6.8 to 7.6~\kms.
Contours begin at, and are in increments of, 0.12~K~km~s\e\
(the dotted contour is at -0.12~K~km~s\e).
The beam size is $14\farcs 8\times 6\farcs 6$ at a position angle
of $-1^\circ$, and is indicated in the lower right corner.}
\label{fig:bima}
\end{figure}
\clearpage

\begin{figure}[htpb]
\centerline{\psfig{figure=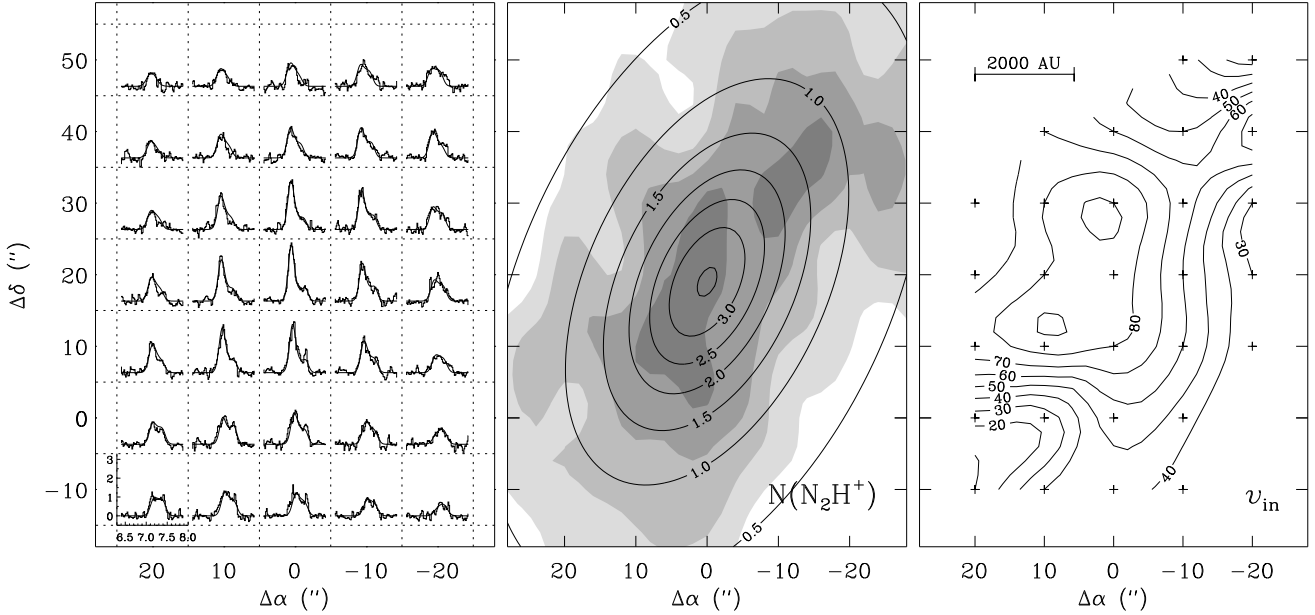,height=5in,angle=90,silent=1}}
\caption{Spectra and model fits for the combined BIMA-IRAM map.
Coordinates are shown offset from $\alpha(2000)=5^{\rm h}04^{\rm m}16^{\rm s}6,
\delta(2000)=25^\circ 10'47\farcs 8$.
The left panel shows observed (heavy histogram) and model (light smooth curve)
spectra on a grid with $10''$ spacing along each axis. The velocity and
temperature axis for each spectrum is illustrated in the lower left corner
and runs from 6.3 to 8.0~\kms in velocity and $-0.5$ to 3.3~K in temperature.
The center panel shows contours of the total model \n2hp\ column density
(front and rear layers, all hyperfine components)
in units of $10^{13}$~cm\ee, overlaid on a grayscale of the
integrated map of \n2hp(101-012) emission from the combined map
(lowest level and increment 0.2~\Kkms).
The right panel plots the velocity difference between the front and
rear layers, \vin, in units of m~s\e. The crosses indicate
those points for which there is sufficient optical depth and signal-to-noise
to determine \vin\ reliably.}
\label{fig:model}
\end{figure}
\clearpage

\begin{table}
\begin{center}
TABLE 1\\
Model Parameters\\
\vskip 2mm
\begin{tabular}{lr}
\hline\\[-2mm]
Parameter  &
Value      \\[2mm]
\hline \hline \\[-3mm]
$x_{\rm pk}$       &  $5^{\rm h}04^{\rm m}16^{\rm s}6$  \nl
$y_{\rm pk}$       &  $25^\circ 11'06\farcs 8$          \nl
$\theta$           &  $25^\circ$                        \nl
$T_{\rm k}$        &  12.0~K                            \nl
$\sigma_{\rm NT}$  &  0.085~\kms                        \nl
$X$                &  $1\times 10^{-9}$                 \nl
$n_{\rm pk,f}$     &  $1\times 10^4$~\cc                \nl
$\Delta z_f$       &  42000~AU                          \nl
$n_{\rm pk,r}$     &  $4\times 10^5$~\cc                \nl
$\Delta z_r$       &  4800~AU                           \nl
$\Delta x$         &  2700~AU                           \nl
$\Delta y$         &  1500~AU                           \nl
$v_0$              &  7.235~\kms                        \nl
$dv/dx$            &  3.8~\kms~pc\e                     \\[2mm]
\hline\\[-2mm]
\end{tabular}
\end{center}
\label{tab:parms}
\end{table}


\begin{references}
\baselineskip=10pt

\reference{AWM96}  Andr\'{e}, P., Ward-Thompson, D., \& Motte, F. 1996
                   \aap, 314, 625

\reference{A95}    Anglada, G. 1995, Rev. Mex. A. \& A., 1, 67

\reference{BM94}   Basu, S., \& Mouschovias, T.C. 1994, \apj, 432, 720

\reference{CMT95}  Caselli, P., Myers P.C., \& Thaddeus, P.
                   1995, \apj, 455, L77

\reference{CM95}   Ciolek, G.E., \& Mouschovias, T.C. 1995, \apj, 454, 194

\reference{GGDB97} Gueth, F., Guilloteau, S., Dutrey, A., \& Bachiller, R.
                   1997, \aap, 323, 943

\reference{L69}    Larson, R.B. 1969, \mnras, 145, 271

\reference{L98}    Li, Z.-Y. 1998, \apj, 493, 230

\reference{LS96}   Li, Z.-Y., \& Shu, F.H. 1996, \apj, 472, 211

\reference{ML98}   Myers, P.C., \& Lazarian, A. 1998, \apj, 507, L157

\reference{My+96}  Myers, P.C., Mardones, D., Tafalla, M., Williams, J.P.,
                   Wilner, D.J. 1996, \apj, 465, L133 (M96)

\reference{T++98}  Tafalla, M., Mardones, D., Myers, P.C., Caselli, P.,
                   Bachiller, R., \& Benson, P.J. 1998, \apj, 504, 900 (T98)

\reference{SAL87}  Shu, F.H., Adams, F.C., \& Lizano, S. 1987, A.R.A\&A., 25, 23

\reference{S77}    Shu, F.H. 1977, \apj, 214, 488

\reference{Z95}    Zhou S. 1995, \apj, 442, 685

\end{references}
\end{document}